\documentclass[a4paper,11pt]{article}
\usepackage{jcappub}

\usepackage{latexsym}
\usepackage{graphicx}
\usepackage{subfigure}
\usepackage{cases}
\usepackage{hyperref}
\usepackage{amssymb}
\usepackage{bm}
\usepackage{natbib}
\usepackage{amsmath}

\newcommand{\be}{\begin{equation}}
\newcommand{\ee}{\end{equation}}
\newcommand{\ba}{\begin{eqnarray}}
\newcommand{\ea}{\end{eqnarray}}

\newcommand{\bk}{{\bm k}}
\newcommand{\bx}{{\bm x}}

\newcommand{\bs}{{{S} ({\bm k}, \eta)}}

\title{CMB Distortions from Damping of Acoustic Waves Produced by Cosmic Strings}

\author{Hiroyuki Tashiro,}
\emailAdd{Hiroyuki.Tashiro@asu.edu}
\author{Eray Sabancilar}
\emailAdd{Eray.Sabancilar@asu.edu}
\author{and Tanmay Vachaspati}
\emailAdd{tvachasp@asu.edu}

\affiliation{Physics Department, Arizona State University, Tempe, Arizona 85287, USA.}

\abstract{
 We study diffusion damping of acoustic waves in the photon-baryon fluid
due to cosmic strings, and calculate the induced $\mu$- and $y$-type
spectral distortions of the cosmic microwave background. For cosmic
strings with tension within current bounds, their contribution to the
spectral distortions is subdominant compared to the distortions from
primordial density perturbations.}

\begin{document}
\maketitle
\flushbottom

\section{Introduction}

Spectral distortions of the cosmic microwave background (CMB) are an important
tool to study the thermal history of the universe at epochs earlier than 
recombination \cite{Zeldovich:1969ff,Sunyaev:1970er}. COBE-FIRAS measurements
of the spectrum of the CMB show a blackbody spectrum to high accuracy, and
result in upper bounds on the spectral distortion parameters,
$|\mu| < 9 \times 10^{-5}$ and $y < 1.5 \times 10^{-5}$
\cite{Mather94,Fixsen:1996nj}. Proposed missions such as PIXIE \cite{Kogut:2011xw}, 
will be able to detect distortion parameters, $\mu$ and $y$, up to levels of 
$10^{-8}$ with 5-$\sigma$ significance, spurring theoretical investigation
of early universe models and their predictions for spectral distortions \cite{Chluba:2011hw,Chluba:2012gq,Tashiro:2012nb,Dent:2012ne,Chluba:2012we,Khatri:2012rt,Pajer:2012dw,Khatri:2012tw}.  

There are various physical mechanisms that can induce spectral distortions, 
including dissipation of acoustic waves in the photon-baryon fluid created by the
primordial density perturbations via Silk damping
\cite{1970Ap&SS...9..368S,1991ApJ...371...14D,Hu:1994bz}, decaying particles
\cite{Hu:1993gc}, electromagnetic radiation from superconducting strings
\cite{Tashiro:2012nb}, Hawking radiation from primordial black holes
\cite{Tashiro:2008sf} and diffusion of magnetic fields
\cite{Jedamzik:1999bm}. Damping of acoustic waves produced by primordial
density perturbations have been studied extensively \cite{Chluba:2011hw,
Chluba:2012gq, Dent:2012ne}. In this paper, we study cosmic strings as an
additional source of acoustic waves, and calculate the resulting CMB distortions 
due to the diffusion damping of these acoustic waves. 

Cosmic strings are predicted in grand unified theories and in string theory (see reviews \cite{VilenkinBook,Polchinski:2004ia,Copeland09,Copeland11}). They can produce various effects, most of which are gravitational, such as CMB anisotropies \cite{Kaiser:1984iv,Allen:1997ag,Albrecht:1997nt,Pogosian:2004ny,Wyman:2005tu,Bevis:2006mj,Bevis:2007gh,Fraisse:2007nu,Pogosian:2008am,Bevis:2010gj,Battye:2010xz,Urrestilla:2011gr,Dvorkin:2011aj,Ade:2013xla}, non-Gaussianity \cite{Hindmarsh:2009qk,Hindmarsh:2009es,Regan:2009hv}, B-mode polarization \cite{Seljak:1997ii,Bevis:2007qz,Pogosian:2007gi,Avgoustidis:2011ax} and gravitational waves \cite{Vachaspati:1984gt,Damour:2000wa,Damour:2001bk,Damour:2004kw,pulsar,Olmez,Sanidas:2012ee,Dufaux12}. The dimensionless parameter that characterizes the strength of gravitational interactions of strings with matter is its tension in Planck units, $G \mu_s$, where $G$ is Newton's constant and $\mu_s$ is the mass per unit length (or tension) of the string. The current limit on the string tension obtained from CMB temperature anisotropies using WMAP \cite{Komatsu:2010fb} and SPT data \cite{Keisler:2011aw} is $G\mu_s \lesssim 1.7\times 10^{-7}$ \cite{Dvorkin:2011aj}, and $G\mu_s \lesssim 1.5 \times 10^{-7}$ \cite{Ade:2013xla} from the recent Planck data \cite{Ade:2013lta}. 

In addition to these effects, cosmic strings induce CMB spectral distortions since they are
relativistic, massive objects, and they stir up the primordial fluid by their
time dependent gravitational potential to generate acoustic waves. This is
a continuous process from the time that the cosmic strings were produced.
Acoustic waves dissipated due to Silk damping in the redshift range $10^6 > z > 5\times 10^4$ produce $\mu-$distortions of the CMB; waves dissipated in the
range $5\times 10^4 > z > 1100$ produce $y-$distortions.

The plan of the paper is as follows. In Sec.~\ref{sec:acoustic}, we solve the continuity and Euler equations for the baryon-photon fluid with a source term, and obtain the acoustic wave energy density due to any source. In Secs.~\ref{sec:longstrings} and \ref{sec:loops}, we calculate CMB distortions from long cosmic strings and loops, respectively. We conclude in Sec.~\ref{sec:conclusions}. We use natural units ($\hbar = c=1$), and assume $\Lambda$CDM cosmology with WMAP values of the cosmological parameters \cite{Komatsu:2010fb}.


\section{Acoustic waves produced by cosmic strings}
\label{sec:acoustic}

The production and evolution of acoustic waves is obtained by solving the 
continuity and Euler equations for the longitudinal (scalar) modes for the 
photon and baryon fluids in Fourier space. We follow the treatment in Ref.~\cite{Hu:1997hp} 
with the inclusion of a source term that describes the effect of strings. The equations for comoving
mode number $k$ are
\begin{eqnarray}
 \dot{ \delta}_{\gamma}
     &=& -{4 \over 3} k v_\gamma ,  
     \label{cont_gamma}
    \\
\dot{v}_{\gamma} &=&
{k \over 4} \delta_\gamma - {k \over 6} \pi_\gamma
- \dot{\tau} ( v_\gamma-v_b) + \bs,
\label{euler_gamma}
\\
\dot{\delta}_b &=& -kv_b,
 \\ 
\dot{v}_b &=& -{\dot{a}\over a}v_b
+ {4\bar\rho_\gamma \over 3\bar\rho_b}
 \dot{\tau}(v_\gamma-v_b) +\bs.
\label{euler_b}
\end{eqnarray}
The subscripts $\gamma$ and $b$ correspond to photon and baryon fluids, 
$\bar{\rho} _{i}$ is the background energy density ($i=\gamma,b$),
$\delta_i$ is the density contrast, $v_i \equiv \hat \bk \cdot {\bm v_i}$ and
$\pi_i \equiv \hat \bk \cdot  {\bm \pi_i}$ are the longitudinal modes of the
velocity and the anisotropic stress, 
$\bs \equiv \hat \bk \cdot  {\bm S}(\bk, \eta)$ is the longitudinal mode of the
source term, $a$ is the scale factor, the overdot represents differentiation 
with respect to conformal time $\eta$, $\dot{\tau} = a n_e \sigma_T$ with
the Thomson scattering cross section, $\sigma_T$, and the electron number
density, $n_e$. We are interested in the cosmic string contribution to spectral
distortions, and so we will use vanishing density perturbations and fluid velocities as initial conditions.

The photon and baryon fluids are tightly coupled on the scales of interest
since $k/{\dot \tau} \ll 1$. Then we expand the difference in velocities
to second order in $k/{\dot \tau}$,
\begin{equation}
v_\gamma-v_b= (k/\dot \tau)f(\tau )+  (k/\dot \tau)^2 g(\tau),
\end{equation}
where $f(\tau) $ and $g (\tau)$ are functions that we will determine. 
We will also ignore terms involving ${\dot a}/a$ since the
damping process is much faster than the Hubble expansion.

To first order, $v_b=v_\gamma -(k / \dot \tau ) f$, and Eqs.~(\ref{euler_gamma}) 
and (\ref{euler_b}) give
\begin{eqnarray}
\dot{v}_{\gamma} &=&
{k \over 4} \delta_\gamma 
- k f + \bs,
\label{euler_gamma_f}
\\
\dot{v}_\gamma &=&
{k  \over R}f 
+\bs,
\end{eqnarray}
where $R= 3\bar\rho_b  / 4\bar\rho_\gamma \propto a$.
To lowest order in $ v_\gamma - v_b$, we drop the anisotropic stress term because 
\begin{equation}
 \pi_\gamma = {96 \over 45} {k \over \dot \tau } v_\gamma,
  \label{190402_22Oct12}
\end{equation}
in the tight coupling approximation.
Eliminating $f$ from both equations, we obtain
\begin{equation}
 \dot{v}_{\gamma} =
{k \over 4 (1+R)} \delta_\gamma + \bs.
\label{euler_1st}
\end{equation}
Comparing the result with Eq.~(\ref{euler_gamma_f}), we get
\begin{equation}
 f= {R \over 4 (1+R) }\delta_\gamma.
\label{funcition_f}
\end{equation}
According to Eqs.~(\ref{cont_gamma}) and (\ref{euler_1st}), we have
\begin{equation}
 \ddot \delta_\gamma = - {k^2 \over 3 (1+R)} \delta_\gamma
  -{4 \over 3} k \bs,
\end{equation}
which shows that the source term will generate acoustic oscillations that
propagate at the sound velocity $c_s = 1/\sqrt{3 (1+R)}$.

To next order, we include the anisotropic stress term (\ref{190402_22Oct12}) and obtain
\begin{eqnarray}
\dot{v}_{\gamma} &=&
{k \over 4 } \delta_\gamma 
-{16 \over 45} {k^2 \over \dot \tau} v_\gamma
- k \left[ {R \over 4(1+R)} \delta_\gamma +
{k \over \dot \tau }g \right] + S,~~~~  \\
\dot{v}_\gamma
 &=&  {k \over \dot \tau} {R \over 4 (1+R) } \dot{\delta}_\gamma +
{k  \over R}  \left[  {R \over 4 (1+R) } \delta_\gamma
	       + {k \over \dot \tau }g \right] + S.~~~~ 
\end{eqnarray}
The difference of these equations yields $g$, and then, using
Eq.~(\ref{cont_gamma}), we get
\be
 \dot{v}_{\gamma} = {k \over 4 (1+R)} \delta_\gamma 
                       + \frac{3}{2} k D \dot \delta_\gamma+\bs,
\label{euler_2nd}
\ee
where
\be
 D= \frac{1}{6 \dot \tau (1+R)} \left[ {R^2 \over 1+R} +{16 \over 15}
  \right ].
\ee
Using Eqs.~(\ref{cont_gamma}) and (\ref{euler_2nd}), we finally obtain
\begin{equation}
\ddot \delta_\gamma = 
 - {k^2 c_s^2} \delta_\gamma - 2 k^2 D \dot \delta_\gamma -{4 \over 3} k \bs.
\label{eq:wave_tight}
\end{equation}

An approximate analytical solution of Eq.~(\ref{eq:wave_tight}) is
\be
 \delta_\gamma (\bk, \eta) \approx {4 \over 3} e^{-k^2 / k_D^2} 
\biggl [ \cos(k r_s) \int^\eta _{\eta_i} \frac{d \eta'}{c_s} ~ {S}({\bk, \eta'})\sin(k r_s' ) - \sin(k r_s) \int ^\eta _{\eta_i} \frac{d \eta'}{c_s}  ~  {S}({\bk, \eta'}) \cos(k r_s' )
 \biggr ] ,
 \label{apxsoln}
\ee
where the sound horizon $r_s$, and the Silk damping scale $1/k_D$ \cite{Silk:1967kq} are
\ba 
r_s' &=& r_s(\eta_i, \eta') =  \int _{\eta_i} ^{\eta'} d\eta~ c_s, \\
~k_D^{-2} &=& \int _{\eta_i} ^{\eta'} d\eta~ D.
\ea
Note that the solution given in Eq.~(\ref{apxsoln}) satisfies Eq.~(\ref{eq:wave_tight}) 
in the tight coupling approximation i.e., we have ignored terms of
order $(k/\dot{\tau})^2$ and higher. Since the exponential damping term $e^{-k^2 / k_D^2}$ depends on $\eta$, it should actually be included inside the integral over $\eta'$ \cite{Battye:1996dg}. However,  we have ignored terms containing $\dot {c}_s$, $\dot R$ and $H = \dot{a}/a$, since the rate of damping of acoustic 
waves is a lot faster than the Hubble expansion rate. Therefore, we have pulled out the exponential damping term. (We have checked that it only introduces an $O(1)$ factor in the final result). The expression in Eq.~(\ref{apxsoln}) applies to a single source and, by linear superposition, to a collection of individual sources. 

The energy density of acoustic (``sound'') waves is given by
\cite{Chluba:2012gq, Khatri:2012rt}
\begin{equation}
 \rho_{\rm s} = {3 \over 8}\bar \rho (1+3 c_s^2) \langle |\delta_\gamma (\bx, \eta)|^2 \rangle
  \approx  {3 \over 4}\bar \rho \langle |\delta_\gamma (\bx,\eta) |^2 \rangle,
\label{rhos}
\end{equation}
where we approximate $c_s^2 \approx 1/3$. The averaged density fluctuation over
a volume $V \rightarrow \infty$ on the right-hand side of Eq.~(\ref{rhos}) is defined 
in terms of Fourier modes by
\be
\langle |\delta_\gamma (\bx ,\eta)|^2 \rangle = 
   \frac{1}{V} \int \frac{d^3 {\bm k}}{(2\pi)^3}~| \delta_\gamma (\bk ,\eta) |^2.
\ee

The rate at which energy density is dissipated from acoustic waves due to Silk
damping is found by substituting Eq.~(\ref{apxsoln}) in Eq.~(\ref{rhos}), and then by taking the time derivative and only retaining the contribution due to Silk damping, i.e., only keeping the term coming from the exponential damping factor as
\be
\dot{Q}(\eta) = \frac{3\bar{\rho}}{2V} \frac{d}{d \eta} (k_{D}^{-2}) \int
\frac{d^3 {\bm k}}{(2\pi)^3} k^2 ~| \delta_\gamma (\bk ,\eta) |^2,
\label{qdot}
\ee
where $\delta (\bk ,\eta)$ is given by Eq.~(\ref{apxsoln}) and $S(\bk,\eta)$ 
is the contribution from all sources. 


\section{CMB distortions due to cosmic strings}
\label{sec:distortion}

In this section, we apply our results for the density contrast given by Eq.~(\ref{apxsoln}) and the rate of dissipated acoustic energy density given by Eq.~(\ref{qdot}) to long cosmic strings and loops, and then calculate the cosmic string contribution to CMB spectral distortions. The spectral $\mu$- and $y$-distortions are given by \cite{Khatri:2012rt}
\ba
\mu &=&  
1.4  \int^{\eta_{\rm freeze}}  _{{\eta_{\mu}} } d\eta 
                                   ~\frac{\dot{Q}(\eta)}{3\bar{\rho}},
\label{mu}
\\
y &=& 
\frac{1}{4} \int_{\eta_{\rm freeze}} ^{\eta_{\rm rec}} d\eta 
                                   ~\frac{\dot{Q}(\eta)}{3\bar{\rho}},
\label{y}
\ea
where $\eta_\mu$, $\eta_{\rm freeze}$ and $\eta_{\rm rec}$ correspond to epochs $z_\mu = 2 \times 10^6$, $z_{\rm freeze} =5 \times 10^4$ and $z_{\rm rec} = 1100$. Note that the factor $1/3$ in Eqs.~(\ref{mu}) and (\ref{y}) comes from the fact
that only 1/3 of the dissipated energy creates the CMB distortions while the
remaining 2/3 is used to raise the average CMB temperature, which cannot be
observed directly \cite{Chluba:2012gq}.


\subsection{Long strings}
\label{sec:longstrings}

We now consider distortions due to long strings. We will adopt the scaling solution 
for the cosmic string network suggested by Nambu-Goto simulations. 
The number density of long strings (in comoving volume) is
\begin{equation}\label{nlong}
 n_{\rm long}(t) = \frac{\kappa}{p}\frac{a^3}{t^3},
\end{equation}
where $a$ is the scale factor, $\kappa = 11$ for the radiation dominated era, and $\kappa=15$ for the matter 
dominated era from the recent simulation results \cite{Shlaer11}. The dependence of the 
density on reconnection probability $p$ is expected to be of the above form \cite{Damour:2004kw,Sakellariadou:2004wq} (however see Ref.~\cite{Avgoustidis:2004zt,Avgoustidis:2005nv}). 
The reconnection probability is unity for ordinary gauge strings, and can be smaller 
than unity for cosmic F- and D-strings, e.g., $10^{-3} \lesssim p \lesssim 1$ \cite{Jackson:2004zg}.

Numerical simulations also show that long strings are not straight, but have wiggles on them. The difference of a wiggly string from a straight one is that its tension, $T$, is not equal to its mass per unit length, $\bar \mu_s$. They satisfy the equation of state given by $T \bar \mu_s = \mu_s^2$, where $\mu_s$ is the bare mass per unit length of the string. The gravitational force due to a wiggly long string is $F \propto (\bar \mu_s - T)/r$ at a distance $r$ from the string. Thus, for a single long wiggly string lying along the $y$-axis  and moving along the $z$-axis with comoving velocity $v_s$, the gravitational force due to the string at the point $\bm x =(x, y, z)$ in comoving coordinates is 
\cite{Vachaspati:1991tt,Vachaspati:1991sy}
\begin{equation}\label{Slongx}
 {\bm S}_{\rm long}^{(i)} ({\bm x}, \eta)= 8 G w \mu_{s} {({\bm x} -v_s \bar \eta \hat
{\bm z} ) \over |{\bm x} -v_s \bar \eta \hat {\bm z}|^2},
\end{equation}
where, $w \equiv (\bar \mu_s - T)/\mu_s$ is the parameter that controls
the wiggleness, $\bar \eta = \eta -\eta_i$ is the elapsed comoving time,
$\hat {\bm z}$ is the unit vector along the $z$-axis, and the
superscript denotes that this is the source term for a single long
string with label $i$. The location, orientation and velocity of the
string will be different for different long strings. We will average
over the locations and orientations, and use the root-mean-square
velocity as seen by simulations.

The Fourier component of the longitudinal mode of the source term, Eq.~(\ref{Slongx}), is
\ba
{S}_{\rm long}^{(i)} ({\bm k}, \eta) &=& \int d^3 {\bm x}~  \hat \bk \cdot {\bm
S}_{\rm long}^{(i)} ({\bm x}, \eta)~ e^{i {\bm k} \cdot {\bm x}}\nonumber
\\
&& \hskip -1.5 cm
= 32 i \pi^2 G \mu_{s} ~ \frac{\delta_D(k_y)}{k_r}~ e^{iv_s \bar \eta k_z}
\left[ 1- J_0(k_r \eta)\right],
\label{Slong}
\ea
where $k^2 = k_y^2 + k_r^2$, $k_r^2 = k_x^2 +k_z^2$, the subscript $D$ in
$\delta_D$ refers to the Dirac delta function, and we assume that the
gravitational force due to the long string does not propagate beyond the
horizon size $\eta$. The total source due to all long strings will be
given by linear superposition
\begin{equation}
S_{\rm long} (\bk,\eta) = \sum_i S_{\rm long}^{(i)} (\bk, \eta).
\end{equation}
The contributions from different long strings will add incoherently, and
Eq.~(\ref{qdot}) gives
\be \dot{Q}_{\rm long} (\eta)
= n_{\rm long} \frac{3\bar{\rho}}{2} \frac{d}{d \eta} (k_{D}^{-2}) \int
\frac{d^3 {\bm k}}{(2\pi)^3} k^2 ~| \delta_\gamma^{(1)} (\bk ,\eta) |^2,
\label{dotQlong}
\ee
where $n_{\rm long}$ is given by Eq.~(\ref{nlong}) and $\delta_\gamma^{(1)}$
is the density fluctuation [see Eq.~(\ref{apxsoln}) for a single source $S^{(1)}$].

We now numerically evaluate the integrals in Eqs.~(\ref{mu}), (\ref{y}) with
${\dot Q}$ from (\ref{dotQlong}) and $S^{(1)}$ from (\ref{Slong}) to get
\ba
\mu \sim  1.  \times 10^{-13}~\frac{w^2}{p} \left({G\mu_{s} \over 10^{-7} } \right)^2,\label{mulong}
\\
y  \sim  2. \times 10^{-14}~ \frac{w^2}{p} \left({G\mu_{s} \over 10^{-7} } \right)^2. \label{ylong}
\label{eq:distort_long}
\ea


\subsection{Loops}
\label{sec:loops}

Having discussed CMB spectral distortions from long strings, we now turn to
cosmic string loops. For a loop of size $L$ and tension $G\mu_{s}$ moving along 
the $z$-axis with comoving velocity $v_s$, the gravitational force at the point
$\bm x =(x, y, z)$ in comoving coordinates is 
\begin{equation}
\label{Sloopx}
{\bm S}_{\rm loop} (\bx, \eta, L) = \frac{G\mu_{s} L}{a} 
  \frac{(\bx-v_s \bar \eta \hat{\bm z})}{|\bx-v_s \bar \eta \hat{\bm z}|^3 }.
\end{equation}
Here $a$ is the scale factor at conformal time $\eta$, $\bar \eta = \eta -\eta_i$ is the elapsed comoving time,  $\eta_i$ is the conformal time at loop production, $L$ is the physical loop size. Note that we model the loop as a point source. This is a good approximation at distances larger than the loop size, where most of the acoustic wave energy  resides. 
 
The Fourier transform of the longitudinal component of the source term
(\ref{Sloopx}) is
\ba
{S}_{\rm loop} ({\bm k}, \eta, L) &=& \int d^3 {\bm x}~  \hat \bk \cdot {\bm
S}_{\rm loop} ({\bm x}, \eta, L)~ e^{i {\bm k} \cdot {\bm x}}\nonumber~~~~~
\\
&=& 4\pi i {G\mu_{s} L \over a} e^{i v_s \bar \eta k_z} \left[ {1 \over k} -
{\sin (k \eta) \over k^2 \eta} \right], ~~~~~
\label{Sloop}
\ea
where $k^2=k_x^2 + k_y^2 +k_z^2$.

The formula for loops analogous to (\ref{dotQlong}) for long strings can be
written as
\be
\dot{Q}_{\rm loop} (\eta) = \frac{3\bar{\rho}}{2} 
  \left [ \int dn_{\rm loop} L^2 \right ]
    \frac{d}{d \eta} (k_{D}^{-2}) \int \frac{d^3 {\bm k}}{(2\pi)^3} k^2 ~
 \biggl | \frac{1}{L} \delta_\gamma^{(1)} (\bk ,\eta) \biggr |^2,
\label{dotQloop}
\ee
where $dn_{\rm loops}$ is the number density of loops of length between
$L$ and $L+dL$. Note that $\delta_\gamma (\bk, \eta) \propto L$ and so
the last factor in (\ref{dotQloop}) is independent of $L$. 

The number density of loops has not been completely resolved\footnote{See
Ref.~\cite{Shlaer11} and references therein for a detailed discussion of
the subject.}. However, substantial evidence for the scaling of loops have been found in Refs.~\cite{Ringeval07,Vanchurin06,Olum07}, and further confirmed in recent simulations \cite{Shlaer11}. In what follows, we
adopt the results of the Nambu-Goto simulation results from
Ref.~\cite{Shlaer11}, where it has been found that a significant fraction of
the energy density that goes into loop production peaks at loops of initial
size $L_i \sim \alpha t_i$, where $\alpha \sim 0.1$ and $t_i$ is the cosmic
time at formation. There is a distribution of loop velocities with an average
of $v_{i} \sim 0.3$ from the same simulations. The number density of cosmic
string loops of length $L$ as a function of time $t$ (in comoving volume) is
\cite{VilenkinBook} 
\begin{numcases}
{dn_{\rm loop}(L,t) = \frac{a^3\kappa_{L}}{p}}  \frac{dL}{t^{3/2} (L + \Gamma
G\mu_s t)^{5/2}}~, & $t < t_{\rm eq}$,\nonumber~~~~~
\\
\frac{ C_L dL}{t^2 (L+\Gamma G\mu_s t)^2}~, & $t > t_{\rm eq}$,~~~~~
\label{dnloop}
\end{numcases}
where $\kappa_{L} \sim 2$ \cite{Shlaer11}, $p\leqslant 1$ is the reconnection
probability, $C_L$ is $\sqrt{{t_{\rm eq}}/(L+\Gamma G\mu_s t)}$ for $L< \alpha
t_{\rm eq}$, and $1$ for $L>\alpha t_{\rm eq}$. In our calculation, we have taken the exponent in Eq.~(\ref{dnloop}) to be $5/2$, however, a slightly different value, namely $2.6$ has been suggested in Ref.~\cite{Lorenz:2010sm} with $\kappa \sim 0.1$.  Our final results for spectral distortions will be reduced by $\sim 0.1$ with the distribution of
Ref.~\cite{Lorenz:2010sm}. 

Note that the distribution given by Eq.~(\ref{dnloop}) takes into account the
energy losses of loops due to gravitational radiation, i.e., as a loop emits
gravitational waves, its length decreases as $L(t) \approx L(t_{i}) -
\Gamma G \mu_{s} t$, where its lifetime is $ \sim L/(\Gamma G \mu_{s})$ with
$\Gamma \sim 50$. 

The loop distribution in Eq.~(\ref{dnloop}) yields
\begin{numcases}
{\int dn_{\rm loop} L^2 \sim \frac{2 a^3\kappa_{L} \sqrt{\alpha}}{p}} 
\sqrt{t_{\rm eq}}/t^{3/2}~,   & $t < t_{\rm eq}$,\nonumber~~~~~~
\\
t_{\rm eq}/t^{2} + \sqrt{\alpha}/t~,  & $t > t_{\rm eq}$,~~~~~~~~
\end{numcases}
and numerical evaluation of the last term in (\ref{dotQloop}) finally gives us
\ba
\mu \sim  2. \times 10^{-13}~ \frac{1}{p} \left({G\mu_{s} \over 10^{-7} } \right)^2, \label{muloops}
 \\
y  \sim  6.  \times 10^{-14}~\frac{1}{p} \left({G\mu_{s} \over 10^{-7} } \right)^2. \label{yloops}
\ea
The COBE-FIRAS measurements of the CMB spectrum constrain
$|\mu| < 9 \times 10^{-5}$ and $|y| < 1.5 \times 10^{-5}$ \cite{Fixsen:1996nj}, and the detection
limits by PIXIE are $\mu > 5 \times 10^{-8}$ and $y> 10^{-8}$ \cite{Kogut:2011xw}.
Therefore, the distortions due to long cosmic strings [Eqs.~(\ref{mulong}) and (\ref{ylong})] and loops [Eqs.~(\ref{muloops}) and Eq.~(\ref{yloops})] are too small to be detected by PIXIE for strings with tension below current bounds, $G\mu_s \lesssim 10^{-7}$.  


\section{Conclusions}
\label{sec:conclusions}

In this paper, we have investigated the evolution of acoustic waves
produced by long cosmic strings and loops in the photon-baryon fluid. Both
long strings and loops can create acoustic waves gravitationally, and
the amplitude is proportional to the string tension, $G \mu_{s}$. 
We have also evaluated the CMB distortions due to the dissipation of
the acoustic waves. 
However, primordial perturbations also produce acoustic waves and
their dissipation is guaranteed to produce CMB
distortion. The acoustic waves from primordial perturbations, with WMAP
cosmological parameters, can produce distortions of order, $\mu \sim 10^{-8}$
and $y\sim 10^{-8}$.  Therefore the distortions due to long cosmic strings are
completely overwhelmed by those produced by the primordial acoustic waves.
On the other hand, long cosmic strings with $G \mu_s \sim 10^{-7}$ can
produce CMB anisotropies with amplitude roughly one hundredth of that
due to primordial acoustic waves at the scale of the sound horizon at the
recombination epoch \cite{Dvorkin:2011aj,Ade:2013xla}.  Considering this fact with 
Eq.~(\ref{eq:distort_long}),
the obtained CMB distortions due to long strings seem to be much smaller than 
what we might have naively expected. The reason is that the amplitude of the
acoustic waves produced by long strings have a highly red spectrum
(spectral index is equal to $-4$)
below the cosmological horizon scale. Therefore, the amplitude of the 
acoustic waves around the Silk scale is much smaller than that of the primordial
acoustic waves. We find that the acoustic wave amplitude due to cosmic string loops
is peaked at the scale $1/k \sim \alpha t/a$. The amplitude is roughly
proportional to $k^2$ below the peak scale.  
Thus, the acoustic waves do not have large enough amplitude around the
Silk scale to create detectable CMB distortions. However, we would like to note that if cosmic strings are superconducting, they can produce detectable CMB distortions \cite{Tashiro:2012nb}. Therefore, CMB distortions can still be useful for studying the effects of cosmic strings.

We would like to emphasize that we only considered fixed loop velocity, $v_s \sim 0.3$, 
which is below the sound speed, $c_s$. In reality there is a distribution of loop velocities \cite{Scherrer:1989ha,Casper:1995ub} with smaller loops being born with larger 
velocities \cite{Copi:2010jw}. 
When $v_s > c_s$, our analysis does not hold, since Eq.~(\ref{eq:wave_tight}) shows that 
there are values of the wavenumber $k$ at which resonance can occur. In addition,
supersonic strings and loops can produce shocks \cite{Stebbins:1987cy,Vachaspati:1991sy}
and these are not included in our analysis. 
However, since cosmic expansion decreases the loop velocity, small loops 
are expected to be supersonic only for a finite period of time. Further, small fast-moving
loops carry only a small fraction of the total energy in loops. So we do not expect supersonic
effects to change our estimates significantly.

\acknowledgments

We would like to thank Eiichiro Komatsu and Matias Zaldarriaga for useful discussions.
This work was supported in part by the Department of Energy, National Science
Foundation Grant No. PHY-0854827 and Cosmology Initiative at Arizona State
University.
 

\bibstyle{aps}


\begin{thebibliography}{999}

\bibitem{Sunyaev:1970er} 
  R.~A.~Sunyaev and Y.~B.~Zeldovich,
 \emph{The Interaction of matter and radiation in the hot model of the universe},
  Astrophys.\ Space Sci.\  {\bf 7} (1970) 20. 

\bibitem{Zeldovich:1969ff} 
  Y.~B.~Zeldovich and R.~A.~Sunyaev,
 \emph{The Interaction of Matter and Radiation in a Hot-Model Universe},
  Astrophys.\ Space Sci.\  {\bf 4} (1969) 301. 

\bibitem{Mather94}
J.~C.~Mather {\it et al.}, Astrophys. J. {\bf 420} (1994) 439. 

\bibitem{Fixsen:1996nj} 
  D.~J.~Fixsen, E.~S.~Cheng, J.~M.~Gales, J.~C.~Mather, R.~A.~Shafer and E.~L.~Wright,
 \emph{The Cosmic Microwave Background spectrum from the full COBE FIRAS data set},
  Astrophys.\ J.\  {\bf 473} (1996) 576 
  \href{http://arxiv.org/abs/astro-ph/9605054}{[arXiv: astro-ph/9605054]}.

\bibitem{Kogut:2011xw} 
  A.~Kogut, D.~J.~Fixsen, D.~T.~Chuss, J.~Dotson, E.~Dwek, M.~Halpern, G.~F.~Hinshaw and S.~M.~Meyer {\it et al.},
 \emph{The Primordial Inflation Explorer (PIXIE): A Nulling Polarimeter for Cosmic Microwave Background Observations},
  JCAP {\bf 1107} (2011) 025 
 \href{http://arxiv.org/abs/1105.2044}{[arXiv:1105.2044 [astro-ph.CO]]}.

\bibitem{Chluba:2011hw} 
  J.~Chluba and R.~A.~Sunyaev,
  \emph{The evolution of CMB spectral distortions in the early Universe},
 Mon. Not. Roy. Astron. Soc, {\bf 419} (2012) 1294  
  \href{http://arxiv.org/abs/1109.6552}{[arXiv:1109.6552 [astro-ph.CO]]}.

\bibitem{Chluba:2012gq} 
  J.~Chluba, R.~Khatri and R.~A.~Sunyaev,
  \emph{CMB at 2x2 order: The dissipation of primordial acoustic waves and the observable part of the associated energy release},
 Mon. Not. Roy. Astron. Soc, {\bf 425} (2012) 1129  
  \href{http://arxiv.org/abs/1202.0057}{[arXiv:1202.0057 [astro-ph.CO]]}.

\bibitem{Tashiro:2012nb} 
  H.~Tashiro, E.~Sabancilar and T.~Vachaspati,
  \emph{CMB Distortions from Superconducting Cosmic Strings},
  Phys.\ Rev.\ D {\bf 85} (2012) 103522
  \href{http://arxiv.org/abs/1202.2474}{[arXiv:1202.2474 [astro-ph.CO]]}.

\bibitem{Dent:2012ne} 
  J.~B.~Dent, D.~A.~Easson and H.~Tashiro,
  \emph{Cosmological constraints from CMB distortion},
  Phys.\ Rev.\ D {\bf 86} (2012) 023514
  \href{http://arxiv.org/abs/1202.6066}{[arXiv:1202.6066 [astro-ph.CO]]}.
 
\bibitem{Chluba:2012we} 
  J.~Chluba, A.~L.~Erickcek and I.~Ben-Dayan,
  \emph{Probing the inflaton: Small-scale power spectrum constraints from measurements of the CMB energy spectrum},
  Astrophys.\ J.\  {\bf 758} (2012) 76
  \href{http://arxiv.org/abs/1203.2681}{[arXiv:1203.2681 [astro-ph.CO]]}.
 
\bibitem{Khatri:2012rt} 
  R.~Khatri, R.~A.~Sunyaev and J.~Chluba,
 \emph{Mixing of blackbodies: entropy production and dissipation of sound waves in the early Universe},
  Astron.\ Astrophys.\  {\bf 543} (2012) A136
  \href{http://arxiv.org/abs/1205.2871}{[arXiv:1205.2871 [astro-ph.CO]]}.
  
\bibitem{Pajer:2012dw} 
  E.~Pajer and M.~Zaldarriaga,
  \emph{A Hydrodynamical Approach to CMB mu-distortions},
  \href{http://arxiv.org/abs/1206.4479}{[arXiv:1206.4479 [astro-ph.CO]]}.
  
\bibitem{Khatri:2012tw} 
  R.~Khatri and R.~A.~Sunyaev,
  \emph{Beyond $y$ and $\mu$: the shape of the CMB spectral distortions in the intermediate epoch, $1.5\times 10^4 < z < 2\times 10^5$},
  JCAP {\bf 1209} (2012) 016
  \href{http://arxiv.org/abs/1207.6654}{[arXiv:1207.6654 [astro-ph.CO]]}.
 
\bibitem[Sunyaev 
\& Zeldovich(1970)]{1970Ap&SS...9..368S}
  R.~A.~Sunyaev and Y.~B.~Zeldovich,
 \emph{The Interaction of matter and radiation in the hot model of the universe},
  Astrophys.\ Space Sci.\  {\bf 9} (1970) 368. 
			
\bibitem[Daly(1991)]{1991ApJ...371...14D}
R.~A.~Daly, Astrophys.\ J.\ {\bf 371} (1991) 14. 
		
\bibitem{Hu:1994bz} 
  W.~Hu, D.~Scott and J.~Silk,
   \emph{Power spectrum constraints from spectral distortions in the cosmic microwave background},
  Astrophys.\ J.\  {\bf 430} (1994) L5 
 \href{http://arxiv.org/abs/astro-ph/9402045}{[arXiv:astro-ph/9402045]}.

\bibitem{Hu:1993gc} 
  W.~Hu and J.~Silk,
 \emph{Thermalization constraints and spectral distortions for massive unstable relic particles},
  Phys.\ Rev.\ Lett.\  {\bf 70} (1993) 2661. 

\bibitem{Tashiro:2008sf} 
  H.~Tashiro and N.~Sugiyama,
 \emph{Constraints on Primordial Black Holes by Distortions of Cosmic Microwave Background},
  Phys.\ Rev.\ D {\bf 78} (2008) 023004 
  \href{http://arxiv.org/abs/0801.3172}{[arXiv:0801.3172 [astro-ph]]}.

\bibitem{Jedamzik:1999bm} 
  K.~Jedamzik, V.~Katalinic and A.~V.~Olinto,
  \emph{A Limit on primordial small scale magnetic fields from CMB distortions},
  Phys.\ Rev.\ Lett.\  {\bf 85} (2000) 700 
  \href{http://arxiv.org/abs/astro-ph/9911100}{[arXiv:astro-ph/9911100]}.
  
  \bibitem{VilenkinBook}
A.~Vilenkin and E.P.S.~Shellard, Cosmic Strings and Other Topological Defects, Cambridge University Press (1994).

\bibitem{Polchinski:2004ia} 
  J.~Polchinski,
  \emph{Introduction to cosmic F- and D-strings}, \href{http://arxiv.org/abs/hep-th/0412244}{[arXiv:hep-th/0412244]}.

\bibitem{Copeland09}
E.J.~Copeland and T.W.B.~Kibble, {\it Cosmic Strings and Superstrings}, Proc. Roy. Soc. Lond. {\bf A 466}, (2010) 623 \href{http://arxiv.org/abs/0911.1345}{[arXiv:hep-th/0911.1345]}.

\bibitem{Copeland11}
E.J.~Copeland, L.~Pogosian and T.~Vachaspati, {\it Seeing String Theory in the Cosmos}, Class. Quant. Grav. {\bf 28},  (2011) 204009 \href{http://arxiv.org/abs/1105.0207}{[arXiv:1105.0207]}.

\bibitem{Kaiser:1984iv} 
  N.~Kaiser and A.~Stebbins,
  {\it Microwave Anisotropy Due to Cosmic Strings},
  Nature {\bf 310}, (1984) 391.

\bibitem{Allen:1997ag} 
  B.~Allen, R.~R.~Caldwell, S.~Dodelson, L.~Knox, E.~P.~S.~Shellard and A.~Stebbins,
 {\it CMB anisotropy induced by cosmic strings on angular scales $>$ approximately 15-minutes},
  Phys.\ Rev.\ Lett.\  {\bf 79}, (1997) 2624 \href{http://arxiv.org/abs/astro-ph/9704160}{[arXiv:astro-ph/9704160]}.

\bibitem{Albrecht:1997nt} 
  A.~Albrecht, R.~A.~Battye and J.~Robinson,
 {\it The Case against scaling defect models of cosmic structure formation},
  Phys.\ Rev.\ Lett.\  {\bf 79}, (1997) 4736 \href{http://arxiv.org/abs/astro-ph/9707129}{[arXiv:astro-ph/9707129]}.

\bibitem{Pogosian:2004ny} 
  L.~Pogosian, M.~C.~Wyman and I.~Wasserman,
  {\it Observational constraints on cosmic strings: Bayesian analysis in a three dimensional parameter space},
  JCAP {\bf 0409}, (2004) 008 \href{http://arxiv.org/abs/astro-ph/0403268}{[arXiv: astro-ph/0403268]}.

\bibitem{Wyman:2005tu} 
  M.~Wyman, L.~Pogosian and I.~Wasserman,
  {\it Bounds on cosmic strings from WMAP and SDSS},
  Phys.\ Rev.\ D {\bf 72}, (2005) 023513
  [Erratum-ibid.\ D {\bf 73}, 089905 (2006)] \href{http://arxiv.org/abs/astro-ph/0503364}{[arXiv:astro-ph/0503364]}.
  
\bibitem{Bevis:2006mj} 
  N.~Bevis, M.~Hindmarsh, M.~Kunz and J.~Urrestilla,
  \emph{CMB power spectrum contribution from cosmic strings using field-evolution simulations of the Abelian Higgs model}
  Phys.\ Rev.\ D {\bf 75}, (2007) 065015
  \href{http://arxiv.org/abs/astro-ph/0605018}{[arXiv:astro-ph/0605018]}.
  
\bibitem{Bevis:2007gh} 
  N.~Bevis, M.~Hindmarsh, M.~Kunz and J.~Urrestilla,
 \emph{Fitting CMB data with cosmic strings and inflation},
  Phys.\ Rev.\ Lett.\  {\bf 100}, (2008)  021301
  \href{http://arxiv.org/abs/astro-ph/0702223}{[arXiv:astro-ph/0702223]}.
  
\bibitem{Fraisse:2007nu} 
  A.~A.~Fraisse, C.~Ringeval, D.~N.~Spergel and F.~R.~Bouchet,
 {\it Small-Angle CMB Temperature Anisotropies Induced by Cosmic Strings},
  Phys.\ Rev.\ D {\bf 78}, (2008) 043535 \href{http://arxiv.org/abs/0708.1162}{[arXiv:0708.1162]}.
  
\bibitem{Pogosian:2008am} 
  L.~Pogosian, S.~H.~H.~Tye, I.~Wasserman and M.~Wyman,
  {\it Cosmic Strings as the Source of Small-Scale Microwave Background Anisotropy},
  JCAP {\bf 0902}, (2009) 013 \href{http://arxiv.org/abs/0804.0810}{[arXiv:0804.0810]}.

\bibitem{Bevis:2010gj} 
  N.~Bevis, M.~Hindmarsh, M.~Kunz and J.~Urrestilla,
 \emph{CMB power spectra from cosmic strings: predictions for the Planck satellite and beyond},
  Phys.\ Rev.\ D {\bf 82}, (2010) 065004
  \href{http://arxiv.org/abs/1005.2663}{[arXiv:1005.2663]}.

\bibitem{Battye:2010xz} 
  R.~Battye and A.~Moss,
 {\it Updated constraints on the cosmic string tension},
  Phys.\ Rev.\ D {\bf 82}, (2010) 023521
  \href{http://arxiv.org/abs/1005.0479}{[arXiv:1005.0479]}.
  
\bibitem{Urrestilla:2011gr} 
  J.~Urrestilla, N.~Bevis, M.~Hindmarsh and M.~Kunz,
 \emph{Cosmic string parameter constraints and model analysis using small scale Cosmic Microwave Background data},
  JCAP {\bf 1112}, (2011) 021
 \href{http://arxiv.org/abs/1108.2730}{[arXiv:1108.2730]}.

\bibitem{Dvorkin:2011aj}
C.~Dvorkin, M.~Wyman and W.~Hu, {\it Cosmic String constraints from WMAP and the South Pole Telescope}, Phys. Rev. {\bf D 84}, (2011) 123519 \href{http://arxiv.org/abs/1109.4947}{[arXiv:1109.4947]}.

\bibitem{Ade:2013xla}
  P.~A.~R.~Ade {\it et al.}  [ Planck Collaboration],
  \emph{Planck 2013 results. XXV. Searches for cosmic strings and other topological defects},
\href{http://arxiv.org/abs/1303.5085}{[arXiv:1303.5085 [astro-ph.CO]]}.

\bibitem{Hindmarsh:2009qk}
  M.~Hindmarsh, C.~Ringeval and T.~Suyama,
  \emph{The CMB temperature bispectrum induced by cosmic strings},
  Phys.\ Rev.\ D {\bf 80} (2009) 083501
 \href{http://arxiv.org/abs/0908.0432}{[arXiv:0908.0432 [astro-ph.CO]]}.
  
\bibitem{Hindmarsh:2009es}
  M.~Hindmarsh, C.~Ringeval and T.~Suyama,
 \emph{The CMB temperature trispectrum of cosmic strings},
  Phys.\ Rev.\ D {\bf 81} (2010) 063505
  \href{http://arxiv.org/abs/0911.1241}{[arXiv:0911.1241 [astro-ph.CO]]}.
  
\bibitem{Regan:2009hv}
  D.~M.~Regan and E.~P.~S.~Shellard,
 \emph{Cosmic String Power Spectrum, Bispectrum and Trispectrum},
  Phys.\ Rev.\ D {\bf 82} (2010) 063527
 \href{http://arxiv.org/abs/0911.2491}{[arXiv:0911.2491 [astro-ph.CO]]}.

\bibitem{Seljak:1997ii} 
  U.~Seljak, U.~-L.~Pen and N.~Turok,
  {\it Polarization of the microwave background in defect models},
  Phys.\ Rev.\ Lett.\  {\bf 79}, (1997) 1615 \href{http://arxiv.org/abs/astro-ph/9704231}{[arXiv:astro-ph/9704231]}.
  
\bibitem{Bevis:2007qz} 
  N.~Bevis, M.~Hindmarsh, M.~Kunz and J.~Urrestilla,
  ``{\it CMB polarization power spectra contributions from a network of cosmic strings}",
  Phys.\ Rev.\ D {\bf 76}, (2007) 043005 \href{http://arxiv.org/abs/0704.3800}{[arXiv:0704.3800]}.

\bibitem{Pogosian:2007gi}
L.~Pogosian and M.~Wyman, {B-modes from Cosmic Strings}, Phys. Rev. {\bf D 77}, (2008) 083509 \href{http://arxiv.org/abs/0711.0747}{[arXiv:0711.0747]}.

\bibitem{Avgoustidis:2011ax}
A.~Avgoustidis {\it et al.}, {\it Constraints on the fundamental string coupling from B-mode experiments}, Phys. 
Rev. Lett. {\bf 107}, (2011) 121301 \href{http://arxiv.org/abs/1105.6198}{[arXiv:1105.6198]}.

\bibitem{Vachaspati:1984gt}
  T.~Vachaspati and A.~Vilenkin,
{\it Gravitational Radiation From Cosmic Strings},
  Phys.\ Rev.\  D {\bf 31}, (1985) 3052.
  
\bibitem{Damour:2000wa}
  T.~Damour and A.~Vilenkin,
 \emph{Gravitational wave bursts from cosmic strings},
  Phys.\ Rev.\ Lett.\  {\bf 85} (2000) 3761
  \href{http://arxiv.org/abs/gr-qc/0004075}{[arXiv:gr-qc/0004075]}.

\bibitem{Damour:2001bk}
  T.~Damour and A.~Vilenkin,
 \emph{Gravitational wave bursts from cusps and kinks on cosmic strings},
  Phys.\ Rev.\ D {\bf 64} (2001) 064008
  \href{http://arxiv.org/abs/gr-qc/0104026}{[arXiv:gr-qc/0104026]}.
  
\bibitem{Damour:2004kw} 
  T.~Damour and A.~Vilenkin,
 \emph{Gravitational radiation from cosmic (super)strings: Bursts, stochastic background, and observational windows},
  Phys.\ Rev.\ D {\bf 71} (2005) 063510
  \href{http://arxiv.org/abs/hep-th/0410222}{[arXiv:hep-th/0410222]}.
  
\bibitem{pulsar}
R. van~Haasteren {\it et al.}, {\it Placing limits on the stochastic gravitational wave backgroung using European pulsar timing array data}, MNRAS {\bf 414 (4)}  (2011) 3117. \href{http://arxiv.org/abs/1103.0576}{[arXiv:1103.0576]}.

\bibitem{Olmez}
S.~Olmez, V.~Mandic and X.~Siemens, {\it Gravitational-wave stochastic background from kinks and cusps on cosmic strings}, Phys. Rev. {\bf D81}, (2010) 104028 \href{http://arxiv.org/abs/1004.0890}{[arXiv:1004.0890]}.

\bibitem{Sanidas:2012ee}
S.A.~Sanidas, R.A.~Battye and B.W.~Stappers, {\it Constraints on cosmic string tension imposed by the limit on the stochastic gravitational wave background from the European Pulsar Timing Array}, Phys.\ Rev.\ D {\bf 85} (2012) 122003 \href{http://arxiv.org/abs/1201.2419}{[arXiv:1201.2419]}.

\bibitem{Dufaux12}
P.~Binetruy, A.~Bohe, C.~Caprini and J.-F.~Dufaux, {\it Cosmological Backgrounds of Gravitational Waves and eLISA/NGO: Phase Transitions, Cosmic Strings and Other Sources}, JCAP {\bf 1206} (2012) 027 \href{http://arxiv.org/abs/1201.0983}{[arXiv:1201.0983]}.

\bibitem{Komatsu:2010fb} 
  E.~Komatsu {\it et al.}  [WMAP Collaboration],
{\it Seven-Year Wilkinson Microwave Anisotropy Probe (WMAP) Observations: Cosmological Interpretation},
  Astrophys.\ J.\ Suppl.\  {\bf 192}, (2011) 18
 \href{http://arxiv.org/abs/1001.4538}{[arXiv:1001.4538 [astro-ph.CO]]}.
  
\bibitem{Keisler:2011aw} 
  R.~Keisler, C.~L.~Reichardt, K.~A.~Aird, B.~A.~Benson, L.~E.~Bleem, J.~E.~Carlstrom, C.~L.~Chang and H.~M.~Cho {\it et al.}, {\it A Measurement of the Damping Tail of the Cosmic Microwave Background Power Spectrum with the South Pole Telescope},
  Astrophys.\ J.\  {\bf 743}, (2011) 28
  \href{http://arxiv.org/abs/1105.3182}{[arXiv:1105.3182 [astro-ph.CO]]}.
  
\bibitem{Ade:2013lta}
  P.A.R.~Ade {\it et al.}  [Planck Collaboration],
  \emph{Planck 2013 results. XVI. Cosmological parameters},
  \href{http://arxiv.org/abs/1303.5076}{[arXiv:1303.5076 [astro-ph.CO]}.


\bibitem{Hu:1997hp} 
  W.~Hu and M.~J.~White,
  \emph{CMB anisotropies: Total angular momentum method},
  Phys.\ Rev.\ D {\bf 56} (1997) 596 
  \href{http://arxiv.org/abs/astro-ph/9702170}{[arXiv:astro-ph/9702170]}.
	
\bibitem{Silk:1967kq} 
  J.~Silk,
  \emph{Cosmic black body radiation and galaxy formation},
  Astrophys.\ J.\  {\bf 151} (1968)  459. 
  
  
\bibitem{Battye:1996dg}
  R.~A.~Battye,
 \emph{Small angle anisotropies in the CMBR from active sources},
  Phys.\ Rev.\ D {\bf 55} (1997) 7361
  \href{http://arxiv.org/abs/astro-ph/9610197}{[astro-ph/9610197]}.
  
 \bibitem{Shlaer11}
J.J. Blanco-Pillado, K. D. Olum and B. Shlaer, 
Phys. Rev. {\bf D 83} (2011)  083514 
\href{http://arxiv.org/abs/1101.5173}{[arXiv:1101.5173 [astro-ph.CO]]}. 

\bibitem{Sakellariadou:2004wq} 
  M.~Sakellariadou,
  \emph{A Note on the evolution of cosmic string/superstring networks},
  JCAP {\bf 0504} (2005) 003
  \href{http://arxiv.org/abs/hep-th/0410234}{[arXiv:hep-th/0410234]}.
  
  

\bibitem{Avgoustidis:2004zt}
  A.~Avgoustidis and E.~P.~S.~Shellard,
 \emph{Cosmic string evolution in higher dimensions},
  Phys.\ Rev.\ D {\bf 71} (2005) 123513
  \href{http://arxiv.org/abs/hep-ph/0410349}{[hep-ph/0410349]}.


\bibitem{Avgoustidis:2005nv}
  A.~Avgoustidis and E.~P.~S.~Shellard,
 \emph{Effect of reconnection probability on cosmic (super)string network density},
  Phys.\ Rev.\ D {\bf 73} (2006) 041301
  \href{http://arxiv.org/abs/astro-ph/0512582}{[astro-ph/0512582]}.


  
\bibitem{Jackson:2004zg} 
  M.~G.~Jackson, N.~T.~Jones and J.~Polchinski,
  \emph{Collisions of cosmic F and D-strings},
  JHEP {\bf 0510} (2005) 013 
  \href{http://arxiv.org/abs/hep-th/0405229}{[arXiv:hep-th/0405229]}.
  
\bibitem{Vachaspati:1991tt} 
  T.~Vachaspati and A.~Vilenkin,
 \emph{Large scale structure from wiggly cosmic strings},
  Phys.\ Rev.\ Lett.\  {\bf 67} (1991) 1057. 

\bibitem{Vachaspati:1991sy} 
  T.~Vachaspati,
 \emph{The Structure of wiggly cosmic string wakes},
  Phys.\ Rev.\ D {\bf 45} (1992) 3487. 
  
 \bibitem{Ringeval07}
C.~Ringeval, M.~Sakellariadou and F.~Bouchet, JCAP {\bf 0702} (2007) 023 
\href{http://arxiv.org/abs/astro-ph/0511646}{[arXiv:astro-ph/0511646]}.

\bibitem{Vanchurin06}
V.~Vanchurin, K.D.~Olum and A.~Vilenkin, Phys. Rev. {\bf D 74} (2006) 063527 
\href{http://arxiv.org/abs/gr-qc/0511159}{[arXiv:gr-qc/0511159]}.

\bibitem{Olum07}
K.D.~Olum and V.~Vanchurin, Phys. Rev. {\bf D 75} (2007) 063521 
\href{http://arxiv.org/abs/astro-ph/0610419}{[arXiv:astro-ph/0610419]}.

\bibitem{Lorenz:2010sm}
L.~Lorenz, C.~Ringeval and M.~Sakellariadou, 
 \emph{Cosmic string loop distribution on all length scales and at any redshift}, 
JCAP {\bf 1010} (2010) 003 
\href{http://arxiv.org/abs/1006.0931}{[arXiv:1006.0931]}.

\bibitem{Scherrer:1989ha} 
  R.~J.~Scherrer and W.~H.~Press,
  \emph{Cosmic String Loop Fragmentation},
  Phys.\ Rev.\ D {\bf 39} (1989) 371. 
  
\bibitem{Casper:1995ub} 
  P.~Casper and B.~Allen,
  \emph{Gravitational radiation from realistic cosmic string loops},
  Phys.\ Rev.\ D {\bf 52} (1995) 4337 
   \href{http://arxiv.org/abs/gr-qc/9505018}{[arXiv:gr-qc/9505018]}.
  
\bibitem{Copi:2010jw} 
  C.~J.~Copi and T.~Vachaspati,
  \emph{Shape of Cosmic String Loops},
  Phys.\ Rev.\ D {\bf 83} (2011) 023529 
  \href{http://arxiv.org/abs/1010.4030}{[arXiv:1010.4030 [hep-th]]}.
  
\bibitem{Stebbins:1987cy} 
  A.~Stebbins, S.~Veeraraghavan, R.~H.~Brandenberger, J.~Silk and N.~Turok,
  \emph{Cosmic String Wakes},
  Astrophys.\ J.\  {\bf 322} (1987) 1. 
	
	
\end{thebibliography}
\end{document}